\documentclass[aps,floatfix,nofootinbib,showpacs,onecolumn,superscriptaddress]{revtex4}
\pdfoutput=1

\usepackage{amsmath}
\usepackage{amssymb}
\usepackage{amsthm}
\usepackage{dcolumn}
\usepackage{epsfig}
\usepackage{graphics}
\usepackage{graphicx}
\usepackage{slashed,epsfig}
\usepackage{longtable}
\usepackage{color}

\definecolor{darkgreen}{rgb}{0,0.5,0}
\definecolor{purple}{rgb}{0.5,0,0.5}
\definecolor{nblue}{rgb}{0.0,0.0,0.50}
\definecolor{scarlet}{rgb}{1.0,0.2,0}
\usepackage[colorlinks=true, pdfstartview=FitV, linkcolor=purple, citecolor= purple, urlcolor=blue]{hyperref}

\newcommand{\nn}{\nonumber}
\newcommand{\beq} {\begin{equation}}
\newcommand{\eeq} {\end{equation}}
\newcommand{\beqa} {\begin{eqnarray}}
\newcommand{\eeqa} {\end{eqnarray}}

\newcommand{\ie}{{\it i.e.}}

\newcommand{\as}{\alpha_s}

\newcommand{\ieps}{i\varepsilon}

\newcommand{\order}[1]{${\cal O}\left(#1 \right)$}
\newcommand{\morder}[1]{{\cal O}\left(#1 \right)}
\newcommand{\eq}[1]{(\ref{#1})}
\newcommand{\fig}[1]{Fig.~\ref{#1}}

\newcommand{\inv}[1]{\frac{1}{#1}}
\newcommand{\ket}[1]{\vert{#1}\rangle}

\newcommand{\ave}[1]{\langle{#1}\rangle}

\newcommand{\acom}[2]{\left\{{#1},{#2}\right\}}

\newcommand{\bs}[1]{\boldsymbol{#1}}

\newcommand{\mA}{\mathcal{A}}

\newcommand{\mD}{\mathcal{D}}

\newcommand{\xv}{{\bs{x}}}
\newcommand{\yv}{{\bs{y}}}
\newcommand{\pv}{{\bs{p}}}
\newcommand{\kv}{{\bs{k}}}

\newcommand{\gv}{\bs{\gamma}}

\newcommand{\halft}{{\textstyle \frac{1}{2}}}

\begin{document}

{\par\raggedleft \texttt{SLAC-PUB-14224}\par}
{\par\raggedleft \texttt{HIP-2010-21/TH}\par}
{\par\raggedleft \texttt{CP3-Origins-2010-34}\par}
\bigskip{}

\title{The $\hbar$ Expansion in Quantum Field Theory}

\author{Stanley~J.~Brodsky} \affiliation{SLAC National Accelerator Laboratory\\
Stanford University, Stanford, California 94309, USA}
\affiliation{CP$^3$-Origins, University of Southern Denmark\\ Campusvej 55, DK-5230 Odense M, Denmark}
\author{Paul Hoyer}
\affiliation{CP$^3$-Origins, University of Southern Denmark\\ Campusvej 55, DK-5230 Odense M, Denmark}
\affiliation{Department of Physics and Helsinki Institute of Physics\\ POB 64, FIN-00014 University of Helsinki, Finland}

\begin{abstract}

We show how expansions in powers of Planck's constant $\hbar = h/2\pi$ can give new insights into perturbative and nonperturbative properties of  quantum field theories. 
Since $\hbar$ is a fundamental parameter, exact Lorentz invariance and gauge invariance are maintained at each order of the expansion.  The physics of the $\hbar$ expansion depends on the scheme; i.e.,  different expansions are obtained depending on which quantities (momenta, couplings and masses) are assumed to be independent of $\hbar$.  We show that  if the coupling and mass parameters  appearing in the Lagrangian density are taken to be independent of $\hbar$, then 
each loop in perturbation theory brings a factor of $\hbar$. In the case of quantum electrodynamics,  this scheme implies that the classical charge $e$, as well as 
the fine structure constant  are linear in  $\hbar$. 
The connection between the number of loops and factors of $\hbar$ is more subtle for bound states since the binding energies and bound-state momenta themselves scale with $\hbar$. The  $\hbar$ expansion allows one to identify equal-time relativistic bound states in QED and QCD which are of lowest order in $\hbar$ and transform dynamically under Lorentz boosts. The possibility to use retarded propagators at the Born level
gives valence-like wave-functions which implicitly describe the sea constituents of the bound states normally present in its Fock state representation.

\end{abstract}


\maketitle

\date{\today}

\section{Introduction}

Planck's constant  $\hbar = h/2\pi$ is the fundamental constant of nature
related to quantum effects \cite{Planck}. The most familiar application of the Planck  constant  is the  commutation relation
$[x_i,p_j] = i \hbar \delta_{ij}$ which limits simultaneous measurements of position and momentum.  More generally, $\hbar$ enters explicitly in the  commutation relations of conjugate operators and fields, thus providing the fundamental basis of quantum field theory. Each order of an expansion in $\hbar $ must obey all the underlying symmetries of the theory. 

The Planck constant has units of action; \ie,  the product of energy and length in units where the velocity of light $c=1$. One commonly assumes units such that $\hbar =1$. However,  it is illuminating to keep the occurrence of powers of $\hbar$ explicit since this  allows one to distinguish quantum versus classical physics. For example, the AdS/CFT prediction for the ratio of shear viscosity to the entropy density of a multi-particle system has the lower limit~\cite{Kovtun:2004de}
${ \eta/ s} \ge {\hbar / 4\pi}.$ The origin of $\hbar$ in this relation can be traced to the assumed quantum form of the entropy of a black hole in a higher dimensional theory.
The empirical observation that $ \eta/ s$ in heavy ion collisions at the Relativistic Heavy Ion Collider \cite{Shuryak:2003xe} is not far from the AdS/CFT prediction thus suggests that the dynamics of high energy central heavy-ion collisions is in the quantum domain.

Physical phenomena are usually expected to follow the laws of classical theory in the (hypothetical) limit $\hbar \to 0$. Surprisingly,  this is true only for a careful choice of $\hbar$-independent quantities (momenta, couplings and masses). A simple illustration is provided by the standard harmonic oscillator in nonrelativistic quantum mechanics where the potential  is $V(x)=\halft m\omega^2 x^2.$
The propagation of a particle from $(t_{i},x_{i})$ to $(t_{f},x_{f})$ is given by the path integral
\beq\label{ho}
\mA(x_{i},x_{f};t_{f}-t_{i}) = \int[\mD x(t)]\exp\left[\frac{i m}{2\hbar}\int_{t_{i}}^{t_{f}}dt ({\dot x}^2-\omega^2 x^2)\right] = \int[\mD\xi(t)]\exp\left[\frac{im}{2}\int_{t_{i}}^{t_{f}}dt ({\dot \xi}^2-\omega^2 \xi^2)\right]
\eeq
In the second equality we have removed the explicit dependence on $\hbar$ by scaling the coordinates as $\xi\equiv x/\sqrt{\hbar}$. (The scaling of the Jacobian is irrelevant for this discussion.)  Remarkably,  the full quantum mechanical structure of the harmonic oscillator model persists as $\hbar \to 0$ when one uses the scaled variables $\xi$, since  the propagation $\xi_{i} \to \xi_{f}$ is independent of $\hbar$, as are the scaled bound-state energies $\epsilon_{n} \equiv E_{n}/\hbar=\omega(n+\halft)$.
Thus there is a domain of positions $x  \propto \sqrt{\hbar}$ and momenta $m\dot x \propto \sqrt{\hbar}$ where the action $S$ is proportional to $\hbar$ and the system stays quantum mechanical even in the $\hbar \to 0$ limit.
On the other hand, the propagation between fixed ($\hbar$-independent) positions $x_{i},x_{f}$ involves large values of $\xi_{i},\xi_{f} \propto 1/\sqrt{\hbar}$, and thus the
transitions between highly excited levels (with $n$  of order $1/\hbar$)
correspond to 
classical dynamics in the $\hbar \to 0$ limit.  

In the case of general relativity, $\hbar$ can be eliminated~\cite{Volovik:2009xs} from the equations of motion, and physical phenomena only depend on dimensionless quantities such as $\alpha_{QED}$.   Although mass cancels out of the equations of motion in classical gravity (due to the equivalence principle), it appears in the Schr\"odinger equation \cite{SakuraiMQM}, but always in the form 
$\tilde m={ m/\hbar}$.   As we shall show, $\tilde m$ is the fundamental mass parameter which appears in the equations of motion for fields when one formulates quantum field theory through the action principle and functional integrals. 

In quantum field theory the $T$-matrix elements of lowest order in $\hbar$ are usually considered to be tree diagrams (Born approximation), with each loop correction  introducing one additional power of $\hbar$ \cite{hbarloops}. However, Donoghue and Holstein and their collaborators \cite{Donoghue:2001qc} have demonstrated that classical physics (of lowest order in $\hbar$) can also emerge from loop diagrams where zero mass quanta appear.
As we shall see, the difference arises from the definition of the $\hbar \to 0$ limit, \ie, the limit depends on which Lagrangian parameters are taken to be independent $\hbar$. 

Born diagrams usually provide a good first approximation to scattering amplitudes in quantum field theory.  However, the  bound-state poles of a scattering amplitude are not present in tree (or any finite number of loop) diagrams, but instead are generated by the divergent  perturbative expansion of the covariant Green's function.  For example, the Schr\"odinger and Dirac bound states,
which arise from tree-level interactions of an electron in an external Coulomb potential, emerge in field theory from the infinite sum of ladder and crossed-ladder 
Feynman diagram contributions to the electron-muon Green's function in the limit where the muon mass is taken to infinity \cite{Brodsky:1971sk}. The binding is caused by loop momenta which are $\propto\hbar$, thus changing the relation between the number of loops and the power of $\hbar$. 

We shall argue that the $\hbar$ expansion can provide a systematic approximation scheme for bound states formed by interactions between particles.
This expansion 
is equally valid for relativistic and non-relativistic dynamics.  We find that the lowest order interaction kernel in $\hbar$ indeed defines a viable ``Born term for bound states''. This Born approximation is insensitive to the $\ieps$ prescription of propagators, which allows a simple Hamiltonian equal-time development in cases where the Coulomb interaction dominates. A hidden Lorentz boost covariance provides a non-trivial test that the approximation correctly includes all contributions of lowest order in $\hbar$.

\section{The $\hbar$ expansion}

There is a general understanding that each loop contribution to quantum field theory amplitudes is associated with one factor of $\hbar$ \cite{hbarloops}. Loop integrals represent quantum fluctuations and classical physics is expected to emerge from the Born term in the $\hbar\to 0$ limit. To fully define this limit one needs to specify the $\hbar$ dependence of the fields, couplings and mass parameters of the action $S$. The choice is not as obvious as it sounds since $\hbar$ appears not only in the path integral measure $\exp(iS/\hbar)$ but (for dimensional reasons) also in the action itself. In this section we establish the scaling with $\hbar$ which makes the loop and $\hbar$ expansions equivalent. In specific problems the energies and momenta may also scale with $\hbar$, as illustrated by the harmonic oscillator above. In the next section we show how the contribution from momenta of \order{\hbar} in the multiple loops of ladder diagrams describes the physics of scattering from a fixed potential and gives rise to Schr\"odinger and Dirac bound states.

We use units where the speed of light $c=\epsilon_0=1.$ Planck's constant $\hbar$ then has dimension of energy $E$ times length $L$;  i.e,  $[\hbar]= E\cdot L$. The dimensions of the fields and other parameters of the action are as usual fixed by the requirement that $S/\hbar$ be dimensionless.

The gluon part of the QCD action,
\beq\label{qcdact}
 -\inv{4} \int d^4x\, (G^{\mu \nu})^2  = -\inv{4} \int d^4x\,  \Big(\partial^\mu A^\nu   - \partial^\nu A^\mu  +i \,\frac{g}{\hbar} \, [A^\mu, A^\nu]\Big)^2
\eeq
sets $[A]= E^{1/2}\cdot L^{-1/2}$. The gluon coupling $g$ has dimension $[g] = E^{1/2}\cdot L^{1/2}$, ensuring that $\as = g^2/4\pi\hbar$ is dimensionless. This requires that $g$ appear with an inverse factor of $\hbar$ in \eq{qcdact}. For the same reason the classical electric charge $e$ and mass $m$ are divided by $\hbar$ in the action ${\cal S}_{SQED}$ of scalar QED,
\beq\label{sqedact}
{\cal S}_{SQED} = \int d^4x\, \big(\big[(\partial^\mu+i\,\frac{e}{\hbar}A^\mu)\phi\big]^\dag \big[(\partial_\mu+i\,\frac{e}{\hbar}A_\mu)\phi\big]-\frac{m^2}{\hbar^2}\phi^\dag\phi\big)
\eeq
The boson field dimensions $[\phi]=[A]= E^{1/2}\cdot L^{-1/2}$ implied by \eq{qcdact} and \eq{sqedact} agree with the dimensions of the corresponding classical fields. Fermion field dimensions are convention dependent since Grassmann numbers have no classical counterparts. While factors of $\hbar$ are rarely displayed in quantum field theory, the convention of Ref. \cite{Sakurai} is to write the electron part of the QED action in the form
\beq\label{qedact}
{\cal S}_{QED} = 
\int d^4x\big[\bar\psi(i\hbar\,\slashed{\partial}-e\slashed{A}-m)\psi\big]
\eeq
with a factor $\hbar$ in the kinetic term. This gives the electron field the dimension $[\psi]=L^{-3/2}$ so that $\psi^\dag\psi$ has the dimension of a probability density. With this convention the classical charge and mass appear in the action without a factor of $\hbar$, and the anti-commutator of the electron field is
\beq\label{fermdim}
\acom{\psi^\dagger(t,\xv)}{\psi(t,\yv)} = \delta^3(\xv-\yv)
\eeq

Insofar as $\hbar$ is regarded as a constant of nature, the factors of $\hbar$ in the action only serve to ensure the proper units of the parameters. However, to define an $\hbar$ expansion we need to specify the dependence on $\hbar$ of all quantities appearing in the action. In analogy to the harmonic oscillator \eq{ho} we take the boson fields to scale as
\beq\label{fieldscale}
\tilde A \equiv A/\sqrt{\hbar},\ \ \ \tilde \phi \equiv \phi/\sqrt{\hbar}
\eeq
with $\tilde A$ and $\tilde \phi$ independent of $\hbar$. Thus $ [\tilde A] = [\tilde \phi ] = L^{-1}$.  Similarly, we take the parameters 
\beq\label{parscale}
\tilde g \equiv \frac{g}{\hbar}\,, \hspace{1cm} \tilde e \equiv \frac{e}{\hbar} \hspace{.5cm} {\rm and} \hspace{.5cm} \tilde m \equiv \frac{m}{\hbar}
\eeq 
to be independent of $\hbar$.  Thus $\tilde m$ has the dimensions of wavenumber $[\tilde m]= L^{-1}$.
In the functional integral measures for the gluon and scalar fields,
\beqa\label{qcdactt}
 -\inv{4\hbar} \int d^4x\, (G^{\mu \nu})^2  &=& -\inv{4} \int d^4x\,  \Big(\partial^\mu\tilde A^\nu   - \partial^\nu\tilde A^\mu  +i \,\tilde{g}\sqrt{\hbar} \, [\tilde A^\mu, \tilde A^\nu]\Big)^2\nn\\
\inv{\hbar}\,{\cal S}_{SQED} &=& \int d^4x\, \big[(\partial^\mu+i\,\tilde e\sqrt{\hbar}\tilde A^\mu)\tilde \phi\big]^\dag \big[(\partial_\mu+i\,\tilde e\sqrt{\hbar}\tilde A_\mu)\tilde\phi\big]-\tilde{m}^2\tilde\phi^\dag\tilde \phi
\eeqa
$\hbar$ then appears exclusively in the combinations $\tilde{g}\sqrt{\hbar}$ and $\tilde{e}\sqrt{\hbar}$. Hence loop corrections of \order{\tilde{g}^2} and \order{\tilde{e}^2} will be of \order{\hbar}.

With the convention of a factor $\hbar$ in the derivative as in \eq{qedact} we take the fermion field to be independent of $\hbar$, while the coupling and mass should scale as in \eq{parscale}. Then
\beq\label{qedactt}
\inv{\hbar}\,{\cal S}_{QED} = 
\int d^4x\Big[\bar\psi(i\,\slashed{\partial}-\tilde e\sqrt{\hbar}\tilde{\slashed{A}}-\tilde m)\psi\Big]
\eeq
again ensuring that each loop is of \order{\hbar}.

The present derivation is equivalent to (albeit perhaps more transparent than) the standard one of Ref. \cite{hbarloops}, which associates a factor $\hbar$ with each propagator, an $\hbar^{-1}$ with each vertex and assumes the parameters appearing in the action to be independent of $\hbar$, \ie, 
$\tilde{g},\tilde{e},\tilde{m}\propto\hbar^0$ 
as in \eq{parscale}. Thus the free gluon propagator $\ave{AA}=\hbar\,\ave{\tilde{A}\tilde{A}}\propto \hbar$ since $\ave{\tilde{A}\tilde{A}}\propto \hbar^0$ at Born level.  Similarly the Born term of 
$\ave{AAA}=\hbar^{3/2}\,\ave{\tilde{A}\tilde{A}\tilde{A}}\propto \tilde{g}\hbar^2$.
The fermion propagator $\ave{\psi\bar\psi}$ is of \order{\hbar^0} with the field normalization implied by \eq{qedact}. Each loop correction adds one power of $\hbar$.\footnote{Reference \cite{hbarloops} uses the normalization convention for fermion fields where there is no factor of $\hbar$ in the kinetic part of the action \eq{qedact}. Then the fermion fields scale similarly to the boson fields \eq{fieldscale}, and identical results are obtained for the powers of $\hbar$ in amplitudes. The normalization convention only affects the Jacobian of the functional integral (and the external lines of the Green function).}

According to \eq{parscale} the equivalence of the loop and $\hbar$ expansions thus, remarkably, requires that the classical couplings $g,\ e$ and masses $m$ {\it divided by $\hbar$} are fixed in the $\hbar\to 0$ limit. For example, the dimensionless fine structure constant,
\beq\label{qedalpha}
\alpha \equiv \frac{e^2}{4\pi\hbar} =\frac{{\tilde e}^2\hbar}{4\pi}
\eeq
is then $\propto\hbar$. The alternative of keeping $e$ and $m$ independent of $\hbar$ was considered in Ref. \cite{Donoghue:2001qc}, where it was found that classical physics can emerge from loops in an amplitude where pairs of massless photons appear in the $t$-channel.

Even though variations of $\hbar$ can not be studied experimentally,  an $\hbar$ expansion is useful as a formal tool which clarifies the structure of theory. The most appropriate expansion scheme depends on the physics at hand. 

In the remaining part of this paper we shall be mostly concerned with bound-state equations which require summing an infinite number of perturbative diagrams. The $\hbar$ expansion  will allow us  to identify a Born term for bound states to which loop corrections may be systematically applied. In particular, we shall make use of the fact that Born terms are insensitive to the $\ieps$ prescription of the propagators, whereas that prescription is crucial for loops.

\section{Application of the $\hbar$ expansion to bound states}

Bound states in quantum field theory can be identified from the divergence of the perturbative series of the $T$-matrix at each bound-state energy. In the familiar case of non-relativistic QED atoms, the divergence arises from ladder diagrams with loop momenta which scale with $\alpha$. The binding energy of hydrogen thus scales as $E_{n}/m_{e} =- \halft \alpha^2/n^2 \propto \hbar^2$. 
Hence 
an $\hbar$ expansion at bound-state poles requires that the external momenta in the $T$-matrix scale with $\hbar$. In such a limit the usual relation between the number of loops and power of $\hbar$ need not hold. The ladder diagrams are in fact sensitive to infrared loop momenta that contribute inverse powers of $\hbar$. The lowest-order contribution in $\alpha$ to the hydrogen atom is also of lowest order in $\hbar$, \ie, it is a Born term for the bound state.
The loop momenta do not scale with $\hbar$ in the higher-order corrections to propagators and vertices,
implying
the usual 
relation between the number of such loops and the power of $\hbar$.

%
\begin{figure}
\includegraphics[width=8cm]{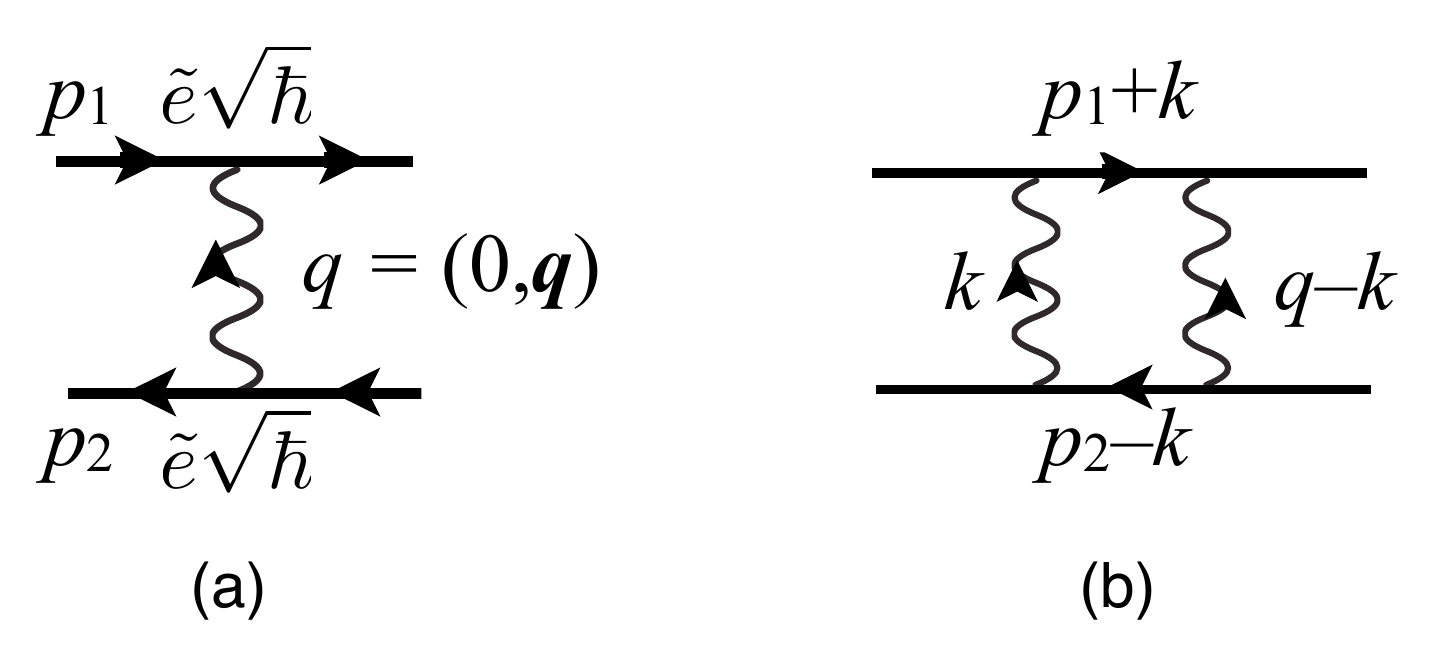}
\caption{\label{Fig1}The first two ladder diagrams contributing to non-relativistic atoms in the limit of small coupling $\alpha$ are shown.}
\end{figure}
%

The bound-state poles of non-relativistic QED atoms are obtained by summing ladder diagrams, the first two of which are shown in \fig{Fig1}. The tree diagram in (a) generally dominates the loop diagram (b) at small coupling strength $\alpha$. However, for external momenta within the range of the atomic wave function the loop diagram is unsuppressed even in the $\alpha \to 0$ limit.  In the center-of-mass system this requires 
$|\pv_{1}|=|\pv_{2}|$ 
to be of $\morder{\alpha m}$ ($m$ being the mass of the constituents) and $|p_{i}^0-m|$ to be of $\morder{\alpha^2 m}$. The restricted range of the loop momentum, $d^4k \propto \alpha^5$ in the bound-state domain, as well as the additional factor of $\alpha$ in the loop diagram, is balanced (relative to the
tree diagram) by its three additional propagators, each of which is of \order{1/\alpha^2}. The situation repeats for all of the higher-order, multi-loop ladder diagrams, allowing the ladder series to diverge and bound-state poles to occur at any value of $\alpha$.

Since $\alpha = {\tilde e}^2\hbar/4\pi \propto \hbar$ the previous argument demonstrates that the loop diagram in \fig{Fig1}(b) is, in the bound-state domain, of the same order also in $\hbar$ as the tree diagram of \fig{Fig1}(a). The loop does not bring an extra factor of $\hbar$ since we consider a limit where the (external and internal) momenta
depend on $\hbar$, 
similarly to the case of the harmonic oscillator \eq{ho}. Repeating the argument for ladder diagrams with more rungs, we may conclude that the Schr\"odinger equation defines the Born term for non-relativistic atoms. The standard higher order corrections in $\alpha$ to the binding energies and wave functions are also of higher order in $\hbar$.

The energy $k^0 \propto \alpha^2 m$ exchanged in the ladder loops is of higher order in $\hbar$ than the 3-momentum $|\kv| \propto \alpha m$. Thus in the Born approximation we may set $k^0=0$ in the photon propagators. This makes the non-relativistic bound-state dynamics equivalent to tree-level scattering from an external, instantaneous potential.

The relativistic Dirac-Coulomb electron bound-states are obtained by summing $e \mu$ scattering diagrams which include all ladder and crossed-ladder photon exchanges \cite{Brodsky:1971sk}. In the limit where the muon mass is large, the sum of crossed and uncrossed photon exchanges gives a $\delta(k^0)$ which suppresses energy exchange and reduces the dynamics to tree-level scattering.  In this sense the relativistic Dirac-Coulomb bound states are also of lowest order in $\hbar$.

\section{Bound states in a static external potential}

The $\ieps$ prescription in propagators is related to the boundary condition in time and is irrelevant at lowest order in $\hbar$. Heuristically, this is seen from the damping factors $\exp(\pm t\delta)$ which give energy denominators $E_{initial}-E_{intermediate} \pm i \epsilon$ where $\epsilon = \delta \hbar$.
In perturbative expansions of the S-matrix the tree diagrams are in fact independent of $\ieps$.
Conversely, the $\ieps$ prescription determines the discontinuities of diagrams which are of higher order in $\hbar$ through the pinching of loop integrals. For Born level bound states this prescription independence allows alternative but equivalent wave functions as we shall next discuss.

Consider first the Dirac-Coulomb bound states in a static external potential. In four-momentum space the interactions with the Coulomb potential $A^0(\kv)$ do not change the energy component $p^0$ of the particle's momentum. Denoting a single Coulomb photon interaction by $K$ the Green's function $G$ of the particle can be expanded as
\beq\label{dseq}
G(p^0,\pv) = S+SKS+SKSKS+\ldots = S+SKG = \frac{R(E,\pv)}{p^0-E}+\ldots
\eeq
In the last equality we displayed the pole contribution of a bound state with energy $E$, whose residue $R(E,\pv)$ is easily seen to satisfy the Dirac equation. Since $K$ preserves $p^0$ the Green's function $G$ is independent of the $\ieps$ prescription at the negative-energy pole of the Dirac propagator $S(p)$. In other words, for $p^0>0$ we have $p^0+E_{p}>0$, where $E_{p} \equiv \sqrt{\pv^2+m^2}$. In particular, the bound-state energy $E$ is the same whether we use a Feynman $S_{F}$ or retarded $S_{R}$ propagator,
\beq\label{frprop}
S_{F/R}(p^0,\pv) \equiv i\frac{\slashed{p}+m}{(p^0-E_{p}+\ieps)(p^0+E_{p}\mp\ieps)}
\eeq

If we time-order the interactions through a Fourier transform $p^0 \to t$ the bound state gives a stationary contribution $G(t,\pv)=\exp(-iEt)R(E,\pv)+\ldots$ However, the Fock state decomposition of its equal-time wave function depends on the choice of propagator \eq{frprop}. In the Feynman propagator $S_{F}(t,\pv)$ the negative energy components move backward in time. This gives rise to ``$Z$''-diagrams describing particle-antiparticle pair fluctuations in the time-ordered interactions with the static potential. The equal-time wave function of a Dirac-Coulomb bound state thus has Fock components with any number of pairs, and its explicit expression is, to the best of our knowledge, not known even in simple cases such as
a $1/r$ potential.

In the case of the retarded propagator $S_{R}$ \eq{frprop}, the negative-energy components move forward in time,
\beqa\label{srt}
S_{R}(t,\pv) = \frac{\theta(t)}{2E_{p}}\left[(E_{p}\gamma^0-\pv\cdot\gv+m)e^{-iE_{p}t} +(E_{p}\gamma^0+\pv\cdot\gv-m)e^{iE_{p}t}\right]
\eeqa
The corresponding time-ordered interactions have no $Z$-contributions, thus only the single bound particle is present at any intermediate time. As a consequence of this, the retarded propagator is local in space at infinitesimal times,
\beq\label{sr0}
\lim_{t\to 0^+}S_{R}(t,\xv) = \gamma^0\delta^3(\xv)
\eeq
By contrast, the right-hand side would be non-local in $\xv$ for the Feynman propagator $S_{F}$. This may be understood by invoking a completeness sum over states at an earlier time: the particle first moves backward in time and then returns to $\xv' \neq \xv$ at $t=0$.

The absence of pair production in retarded propagation as well as the locality property \eq{sr0} allows a one-particle Hamiltonian description as in relativistic quantum mechanics \cite{Hoyer:2009ep}. The bound-state wave functions are then given by the Dirac equation. These single-particle wave functions describe the same bound states, which using Feynman propagators, contain an indefinite number of particle pairs arising from $Z$-diagrams. As we have seen from \eq{dseq}, the bound-state energies $E$ of both pictures agree at lowest order in $\hbar$.

\section{Relativistic bound states in field theory}

The Hamiltonian description of relativistic bound states in an external Coulomb potential can in certain cases be extended to field theory \cite{Hoyer:2009ep}. According to the above arguments the binding energies will at Born level be independent of the $\ieps$ prescription used in the bound particle propagators. This allows the use of the retarded propagator $S_{R}$ \eq{srt} which suppresses $Z$-diagrams and makes the time evolution local as in \eq{sr0}. Coulomb exchange gives an instantaneous potential due to the absence of $\partial_{0}A^0$ terms in the Lagrangian density. Transversely polarized photons propagate in time and thus they generate higher Fock states. Here we only consider cases where Coulomb exchange is dominant.

In $D=1+1$ dimensions only Coulomb exchange contributes in a gauge where $A^1=0$. For an $e^-\mu^+$ state where the electron is at $x_{1}$ and the muon at $x_{2}$
\beq\label{ffstate}
\ket{E,t=0}=\int dx_{1}dx_{2}\,\psi_{e}^\dag(t=0,x_{1})\chi(x_{1},x_{2})\psi_{\mu}(t=0,x_{2})\ket{0}
\eeq
the equation of motion (Gauss' law),
\beq\label{gauss}
-\partial_{x}^2 A^0(x;x_{1},x_{2})= e\left[\delta(x-x_{1})-\delta(x-x_{2})\right]
\eeq
determines the instantaneous Coulomb field,
\beq\label{a0pot}
-A^0(x;x_{1},x_{2})= \halft e\big[\,|x-x_{1}|-|x-x_{2}|\,\big]
\eeq
It is then straightforward \cite{Hoyer:2009ep} to determine the condition on the bound-state wave function $\chi(x_{1},x_{2})$ which ensures a stationary time development,
\beq\label{ffqed}
\gamma^0(-i{\buildrel\rightarrow\over\partial}_{x_{1}}\,\gamma^1 +m_{e})\chi(x_{1},x_{2}) - \chi(x_{1},x_{2})\gamma^0(i{\buildrel\leftarrow\over\partial}_{x_{2}}\,\gamma^1 +m_{\mu}) = [E-V(x_{1}-x_{2})] \chi(x_{1},x_{2})
\eeq
Here the kinetic term of the electron operates on $\chi(x_{1},x_{2})$ from the left, and that of the muon from the right.
The potential following from \eq{a0pot} is
\beq\label{pot}
V(x)=\halft e^2 |x|
\eeq
In $D=1+1$ dimensions the Dirac matrices as well as the wave function $\chi(x_{1},x_{2})$ may be taken to be $2\times 2$ matrices.

Not surprisingly, the bound-state equation \eq{ffqed} has a ``double Dirac'' form, and as such was proposed already by Breit \cite{Breit:1929zz}. Now this equation is seen to provide an approximation of lowest order in $\hbar$ to relativistic bound states,
with the potential \eq{pot} in 1+1 dimensions fixed by QED.

A stringent test that \eq{ffqed} actually represents the exact Born level result is that it is consistent with the Poincar\'e invariance of QED. Translational invariance is explicit, but the boost invariance is hidden (dynamic), since the constituents are at equal time in all frames. It turns out \cite{Hoyer:1985tz} that for the linear potential \eq{pot} (and for no other form of the potential) the bound-state energy indeed has the correct dependence on the CM momentum $P$ of the bound state,
\beq
E=\sqrt{P^2+M^2}
\eeq
The $P$-dependence of the wave function $\chi(x_{1},x_{2})$ is non-trivial. It resembles an ordinary Lorentz transformation, but with $E \to E-V(x)$ (the canonical energy) in the boost parameter. Thus the wave function contracts at a rate which depends on the separation $x$ between the constituents.

The properties of the relativistic bound states defined by \eq{ffqed} merit further study. At large separations between the constituents, where $V(x)\gg E$, the wave function has a constant, non-vanishing density in $x$. This may
reflect the particle pairs which are polarized from the vacuum in a formulation using Feynman propagators.

The two very different but equivalent pictures of bound states that we find here using retarded {\it vs.} Feynman time development may be related to the well-known puzzle of the 
quark model {\it vs.} parton model views of real hadrons: Hadron excitation spectra reflect mainly their valence quark degrees of freedom, even though sea quarks contribute importantly to deep inelastic scattering at low $x_{Bj}$ and low $Q^2$ \cite{cooper}.

A bound-state equation analogous to \eq{ffqed} may be derived in $D=3+1$ dimensions by assuming a non-trivial boundary condition in Gauss' law \eq{gauss}. The homogenous solution for $A^0$ then gives rise to a linear potential. Poincar\'e invariance is respected similarly to the 1+1-dimensional case. It is also possible to derive the analogous Born level meson and baryon bound states in QCD \cite{Hoyer:2009ep}. The $qqq$ potential is gauge covariant and confines the three quarks in a symmetric way. In some respects this resembles the soft-wall AdS/QCD models~\cite{deTeramond:2008ht,Karch:2006pv} which utilize a linear potential in an effective Dirac equation in AdS space.

\section{Conclusions}

The Planck constant $\hbar = h/2\pi$ is a fundamental constant of nature which gives a measure of quantum effects.
In this paper we have shown how expansions in powers of Planck's constant $\hbar $ can give new insights into perturbative and nonperturbative properties of  quantum field theories. 
Since $\hbar$ is a fundamental parameter, exact Lorentz invariance and gauge invariance are maintained at each order of the expansion.  

It is common to set $\hbar=1$ since there is a general belief that the power of $\hbar$ is given by the number of loops in the perturbative expansion \cite{hbarloops}. In the functional integral $\hbar$ appears 
in the measure $\exp(i {\cal S}/\hbar)$, and the action $\cal S$ is usually assumed to be independent of $\hbar$. It is then argued that classical physics emerges in the $\hbar \to 0$ limit since the rapidly varying phase ${\cal S}/\hbar$ selects field configurations for which the action $\cal S$ is stationary. This argument is, however, somewhat oversimplified since there may be field configurations which make ${\cal S} \propto \hbar$. We have illustrated this with the harmonic oscillator, whose full quantum mechanical bound-state spectrum persists in the $\hbar \to 0$ limit.

In fact, the physics of the $\hbar$ expansion depends on the scheme; i.e.,  different expansions are obtained depending on which quantities (momenta, couplings and masses) are assumed to be independent of $\hbar$.
We have shown that  if the coupling and mass parameters  appearing in the Lagrangian density are taken to be independent of $\hbar$, then 
each loop in perturbation theory brings a factor of $\hbar$. In the case of quantum electrodynamics,  this scheme implies that the classical charge $e$, as well as 
the fine structure constant  are linear in  $\hbar$.

We have also noted that for dimensional reasons the QCD coupling $g$ is divided by $\hbar$ in the action \eq{qcdact}. Similarly, the classical charge $e$ and mass $m$ are divided by $\hbar$ in the QED action \eq{sqedact}.

It may seem natural to fix the classical quantities $g$, $e$ and $m$ 
as $\hbar\to 0$, as is done in \cite{Donoghue:2001qc}. However, this scheme introduces a non-trivial dependence on $\hbar$ into the action ${\cal S}$. We also note that the fine structure constant 
$\alpha= e^2 /4\pi \hbar c$ diverges in the $\hbar \to 0$ limit if the classical coupling $e$ is held constant.  Furthermore, it has been demonstrated through explicit examples in this scheme that loops containing massless quanta, such as the gravitational form factors of the electron, can contribute to classical physics in the $\hbar\to 0$ limit \cite{Donoghue:2001qc}. The underlying reason is that an $\hbar$ expansion is not uniquely defined without specifying which quantities (momenta, couplings and masses) are to be regarded as independent of $\hbar$. 

On the other hand, if the parameters of the
action, $\tilde g=g/\hbar$, $\tilde e=e/\hbar$ and $\tilde m=m/\hbar$ 
are taken to be independent of $\hbar$, the fine structure constant $\alpha={\tilde e}^2\hbar/4\pi\propto \hbar$ and the $\hbar$ expansion is equivalent to the perturbative expansion.  For example, in the
$\hbar \to 0$ limit, the loops which define the $\beta$ function vanish, so that the running coupling $\tilde e(\mu^2)$ is constant as in a conformal theory.

The $\hbar$ expansion is particularly non-trivial and illuminating in the case of bound states. We have considered whether one can define a Born term for bound states, which is equivalent to a tree (no loop) approximation when $\tilde g$, $\tilde e$ and  $\tilde m$ are fixed. This would allow an unambiguous and physically motivated approximation scheme for relativistic bound states, maintaining all symmetries of the theory at each order in $\hbar$. In the familiar case of non-relativistic QED atoms the binding energy $E_{bind} \propto \alpha^2 m \propto \hbar^3 \tilde m$ depends on $\hbar$. In order to stay on the bound state pole of a Green's function in the $\hbar \to 0$ limit we must allow the momenta to scale with $\hbar$, which introduces additional sources of $\hbar$. This is why the sum of multi-loop ladder diagrams in perturbation theory reduces to a Born approximation for atomic states.

Born terms are insensitive to the $\ieps$ prescription of perturbative propagators. This is explicitly seen for tree diagrams of the S-matrix, and we  have shown that it holds also for bound states in a Coulomb potential. Using Feynman or retarded propagators does, however, make a consequential difference for the wave functions of the equal-time bound states. Relativistic bound states given by the Dirac equation have an indefinite number of particle pair constituents in their Fock expansion which arise from $Z$-diagrams
due to Feynman propagators. In contrast, if one chooses retarded propagators, the bound state appears to contain just a single particle (of positive or negative energy) whose distribution is given by the standard Dirac wave function.

Our understanding of bound states formed by the mutual interactions of relativistic particles is still very limited. This may be due to the lack of a physically well motivated and manageable first approximation akin to the tree diagrams of scattering amplitudes.  In this paper  we have shown that there is a well-defined Born approximation for relativistic bound states. With retarded propagation one obtains simple and specific bound-state equations. The two-body equations in QED and QCD have a hidden boost invariance which strongly suggests that they include all effects of lowest order in $\hbar$. Just as for the Dirac equation, the resulting valence wave functions implicitly describe the multiple pair constituents generated by Feynman propagation.

\acknowledgements

We thank J. F. Donoghue and B. R. Holstein for a critical reading of our manuscript and a helpful discussion concerning the definition of $\hbar$ expansions. We have also benefitted from discussions with
C. Carlson and S. Peign\'e.

\end{document}